\documentclass{PoS}
\usepackage{slashed}

\title{From $J/\psi$ to LHCb pentaquarks}

\ShortTitle{From $J/\psi$ to LHCb pentaquarks}

\author{\speaker{F. Fernandez}\\
Instituto de F\'isica Fundamental y Matem\'aticas (IUFFyM), Universidad de 
Salamanca, E-37008 Salamanca, Spain \\
E-mail: \email{fdz@usal.es}}

\author{P. G. Ortega\\
Instituto de F\'isica Corpuscular (IFIC), Centro Mixto CSIC-Universidad de 
Valencia, E-46071 Valencia, Spain\\
E-mail: \email{pgortega@ific.uv.es}}

\author{D. R. Entem\\
Instituto de F\'isica Fundamental y Matem\'aticas (IUFFyM), Universidad 
de Salamanca, E-37008 Salamanca, Spain \\
E-mail: \email{entem@usal.es}}

\author{J. Segovia\\
Physik-Department, Technische Universit\"at M\"unchen, 
James-Franck-Str.~1, 85748 Garching, Germany\\
E-mail:\email{jorge.segovia@tum.de}}

\abstract{The two exotic $P_c^+(4380)$ and $P_c^+(4450)$ discovered in $2015$ 
by the LHCb Collaboration, together with the four resonances $X(4140)$, 
$X(4274)$, $X(4500)$ and $X(4700)$, reported in $2016$ by the same 
collaboration, are described in a constituent quark model which has been able 
to explain the properties of charmonium states from the $J/\psi$ to the 
$X(3872)$. Using this model we found a $\bar D\Sigma_c^*$ bound state with 
$J^P=\frac{3}{2}^-$ that may be identified with the $P_c^+(4380)$. In the  
$\bar D^*\Sigma_c$ channel we found three possible candidates for the 
$P_c^+(4450)$ with  $J^P=\frac{1}{2}^-$, $\frac{3}{2}^-$ and $\frac{3}{2}^+$ 
with almost degenerated energies. The $X(4140)$ resonance appears as a cusp in 
the $J/\psi\phi$ channel due to the near coincidence of the 
$D_{s}^{\pm}D_{s}^{\ast\pm}$ and $J/\psi\phi$ mass thresholds. The remaining 
three $X(4274)$, $X(4500)$ and $X(4700)$ resonances appear as conventional 
charmonium states with quantum numbers $3^{3}P_{1}$, $4^{3}P_{0}$ and 
$5^{3}P_{0}$, respectively; and whose masses and widths are slightly modified 
due to their coupling with the corresponding closest meson-meson thresholds.}

\FullConference{VIII International Workshop On Charm Physics\\
		5-9 September, 2016\\
		Bologna, Italy}

\begin{document}


\section{Introduction}

Since the mythical date of November 1974 when the $J/\psi$ particle was 
discovered, two new dates have become important in Charm Physics, enriching the 
field with new and interesting structures. In 2003 Belle Collaboration 
discovered two new states, the $X(3872)$ and the $D_{s0}(2317)$, whereas CLEO 
Collaboration measured the properties of the $D_{s1}(2460)$. All these states 
can hardly be accommodated in the predictions of the naive (but successful) 
quark models. This year represents the beginning of the discovery of a series of 
states collectively called $XYZ$ states whose description is a challenge for 
theorists.

In the last two years the discovery by the LHCb Collaboration of several new 
resonances provoked a lively discussion about their structures 
(see~\cite{Chen2016} for a review). Two resonances, $P_c^+(4380)$ and 
$P_c^+(4450)$, compatibles with a pentaquark state, were observed in 2015 in the 
$J/\psi$ invariant mass spectrum of the $\Lambda^0_b\rightarrow J/\psi K^-p$ 
process~\cite{Aaij:2015tga}. The values of the masses and widths from a fit 
using Breit-Wigner amplitudes are $M_{P_c(4380)}=(4380\pm 8\pm 29)$~MeV/c$^2$, 
$\Gamma_{P_c(4380)}=(205\pm 18\pm 86)$~MeV, $M_{P_c(4450)}=(4449.8\pm 1.7\pm 
2.5)$~MeV/c$^2$ and $\Gamma_{P_c(4450)}=(39\pm 5\pm 19)$~MeV.

Thanks to the large signal yield, the roughly uniform efficiency and the 
relatively low background across the entire $J/\psi \phi$ mass range, the LHCb 
Collaboration~\cite{Aaij:2016iza, Aaij:2016nsc} observed in the $J/\psi \phi$ 
invariant mass of the $B^+\rightarrow J/\psi \phi K^+$ decay the $X(4140)$, 
$X(4274)$, $X(4500)$ and $X(4700)$ resonances. The quantum numbers of the 
$X(4140)$ and $X(4274)$ states are determined to be $J^{PC}=1^{++}$ with 
statistical significance $5.7\sigma$ and $5.8\sigma$, respectively. The 
$X(4500)$ and $X(4700)$ resonances have both $J^{PC}=0^{++}$ with statistical 
significance $4.0\sigma$ and $4.5\sigma$, respectively. 

The $X(4140)$ was measured first by the CDF Collaboration~\cite{Aalt2009}, 
which in an unpublished update of their analysis~\cite{Aalt2011} reported a 
second narrow peak near $4270$ MeV. However, Belle and LHCb Collaborations do 
not found any signal of these resonances in two photon collision and $B^+$ 
decays respectively.

Most of the $XYZ$ resonances appear near open charm meson-meson thresholds and 
are explained as a coupled channel effect of $c\bar c$ structures with the 
meson-meson open channels. If the interaction is strong enough a new meson-meson 
(molecular) state appears besides the $c\bar c$ states, as in the case of the 
$X(3872)$~\cite{Ortega:2010qq} but if the coupling is weak the only effect is 
the renormalization of the mass of the $c\bar c$ states, as in the case of the 
$D_{s0}(2317)$~\cite{Ortega:2016mms}. Sometimes the meson-meson interaction is 
the responsible of forming the resonance state without the coupling to a $c\bar 
c$ state. As all the new resonances mentioned above appear near meson-meson or 
meson-baryon thresholds, we will explore the possibility of describing the new 
states within the same coupled channel scheme. To do that, we use the 
constituent quark model (CQM) presented in Ref.~\cite{Vijande:2004he} and 
updated in Ref.~\cite{Segovia:2008zz} which has been extensively used to 
describe the hadron phenomenology. We use the same set of parameters of 
Ref.~\cite{Segovia:2008zz} without introducing any new one for the calculation. 
This is important, because the modification of the number or value of the 
parameters may modify arbitrarily the meson-meson or meson-baryon interaction 
and, thus, create artificially new states.


\section{The model}

The constituent quark model we use is based on the assumption that the light 
constituent mass appears due to the spontaneous chiral symmetry breaking of QCD 
at some momentum scale. Regardless of the breaking mechanism, the simplest 
Lagrangian which describe this situation must contain chiral fields to 
compensate the mass term and can be expressed as~\cite{Diakonov:2002fq}
\begin{equation}\label{lagrangian}
{\mathcal L}
=\overline{\psi }(i\, {\slashed{\partial}} -M(q^{2})U^{\gamma_{5}})\,\psi 
\end{equation}
where $U^{\gamma _{5}}=\exp (i\pi ^{a}\lambda ^{a}\gamma _{5}/f_{\pi })$, 
$\pi^{a}$ denotes nine pseudoscalar fields $(\eta _{0,}\vec{\pi },K_{i},\eta 
_{8})$ with $i=$1,...,4 and $M(q^2)$ is the constituent mass. This constituent 
quark mass, which vanishes at large momenta and is frozen at low momenta at a 
value around 300 MeV, can be explicitly obtained from the theory but its 
theoretical behavior can be simulated by parametrizing $M(q^{2})=m_{q}F(q^{2})$ 
where $m_{q}\simeq $ 300 MeV, and
\begin{equation}
F(q^{2})=\left[ \frac{{\Lambda}^{2}}{\Lambda ^{2}+q^{2}}
\right] ^{\frac{1}{2}} \, .
\end{equation} 
The cut-off $\Lambda$ fixes the chiral symmetry breaking scale.

The Goldstone boson field matrix $U^{\gamma _{5}}$ can be expanded in terms of 
boson fields,
\begin{equation}
U^{\gamma _{5}}=1+\frac{i}{f_{\pi }}\gamma ^{5}\lambda ^{a}\pi ^{a}-\frac{1}{%
2f_{\pi }^{2}}\pi ^{a}\pi ^{a}+...
\end{equation}
The first term of the expansion generates the constituent quark mass while the
second gives rise to a one-boson exchange interaction between quarks. The
main contribution of the third term comes from the two-pion exchange which
has been simulated by means of a scalar exchange potential.

In the heavy quark sector chiral symmetry is explicitly broken and we do not 
need to introduce additional fields. However the chiral fields introduced above 
provide a natural way to incorporate the pion exchange interaction in the 
molecular dynamics.

The other two main properties of QCD (besides the chiral symmetry breaking) are 
confinement and asymptotic freedom. At present it is still infeasible to 
analytically derive these properties from the QCD Lagrangian, hence we model 
the interaction with a phenomenological confinement and the one-gluon exchange 
potentials, the last one, following De Rujula~\cite{DeRujula:1975qlm}, coming 
from the Lagrangian.

\begin{equation}
\label{Lg}
{\mathcal L}_{gqq}=
i{\sqrt{4\pi\alpha _{s}}}\, \overline{\psi }\gamma _{\mu }G^{\mu
}_c \lambda _{c}\psi  \, ,
\end{equation}
where $\lambda _{c}$ are the SU(3) colour generators and G$^{\mu}_c$ the gluon 
field. 
  
The confinement term, which prevents from having coloured hadrons, can be 
physically interpreted in a picture where the quark and the antiquark are 
linked by a one-dimensional colour flux-tube. The spontaneous creation of 
light-quark pairs may give rise at same scale to a breakup of the colour 
flux-tube. This can be translated into a screened potential, in such a way that 
the potential saturates at the same interquark distance, such as
\begin{equation}
V_{CON}(\vec{r}_{ij})=\{-a_{c}\,(1-e^{-\mu_c\,r_{ij}})+ \Delta\}(\vec{%
\lambda^c}_{i}\cdot \vec{ \lambda^c}_{j})\,
\end{equation}
where $\Delta$ is a global constant to fit the origin of energies. Explicit 
expressions for all these interactions are given in Ref.~\cite{Vijande:2004he}. 
In the same reference all the parameters of the model are detailed, 
additionally adapted for the heavy meson spectra in Ref.~\cite{Segovia:2008zz}.

In order to find the quark-antiquark bound states with this constituent quark 
model, we solve the Schr\"odinger equation using the Gaussian expansion 
method~\cite{Hiyama:2003cu} (GEM), expanding the radial wave function in terms 
of basis functions 
\begin{equation}
R_{\alpha}(r)=\sum_{n=1}^{n_{max}} c_{n}^\alpha \phi^G_{nl}(r),
\end{equation}
where $\alpha$ refers to the channel quantum numbers and $\phi^G_{nl}(r)$ are 
Gaussian trial functions with ranges in geometric progression. This choice is 
useful for optimizing the ranges with a small number of free 
parameters~\cite{Hiyama:2003cu}. In addition, the geometric progression is 
dense at short distances, so that it enables the description of the dynamics 
mediated by short range potentials.

The coefficients, $c_{n}^\alpha$, and the eigenvalue, $E$, are determined from 
the Rayleigh-Ritz variational principle
\begin{equation}
\sum_{n=1}^{n_{max}} \left[\left(T_{n'n}^\alpha-EN_{n'n}^\alpha\right)
c_{n}^\alpha+\sum_{\alpha'}
\ V_{n'n}^{\alpha\alpha'}c_{n}^{\alpha'}=0\right],
\end{equation}
where $T_{n'n}^\alpha$, $N_{n'n}^\alpha$ and $V_{n'n}^{\alpha\alpha'}$ are the 
matrix elements of the kinetic energy, the normalization and the potential, 
respectively. $T_{n'n}^\alpha$ and $N_{n'n}^\alpha$ are diagonal, whereas the
mixing between different channels is given by $V_{n'n}^{\alpha\alpha'}$.

Following Ref.~\cite{Valcarce:2005em}, in order to model the meson-baryon 
system we use a Gaussian form to describe the baryon wave function,
\begin{equation} 
\psi (\vec{p}_i)=\prod_{i=1}^3 \left[ \frac{\alpha_i b^2}{\pi}\right]^{\frac{3}{4}} 
e^{-\frac{b^2 \alpha_i p_i^2}{2}},
\end{equation}
where we take the values $b=0.518\,fm$ and $\alpha_i=1$ for the nucleon wave 
function~\cite{Valcarce:2005em}, and the scaling parameters $\alpha_i$ for 
different flavours are obtained using the prescription of 
Ref.~\cite{Straub:1988gj}.

To describe the coupling between two and four quark configuration, we assume 
now that the hadronic state can be described as 
\begin{equation} 
| \Psi \rangle = \sum_\alpha c_\alpha | \psi_\alpha \rangle
+ \sum_\beta \chi_\beta(P) |\phi_A \phi_B \beta \rangle,
\label{eq:funonda}
\end{equation}
where $|\psi_\alpha\rangle$ are $c\bar{c}$ eigenstates solution of the two-body 
problem, $\phi_{A}$ and $\phi_{B}$ are the two meson states with $\beta$ 
quantum numbers, and $\chi_\beta(P)$ is the relative wave function between the 
two mesons.

Two- and four-quark configurations are coupled using the same transition 
operator that has allowed us to compute the above open-flavour strong decays. 
This is because the coupling between the quark-antiquark and meson-meson 
sectors requires also the creation of a light quark 
pair~\cite{LeYaouanc:1972ae, LeYaouanc:1973xz}. We define the transition 
potential $h_{\beta \alpha}(P)$ within the $^{3}P_{0}$ 
model as~\cite{Kalashnikova:2005ui} 
\begin{equation}
\langle \phi_{A} \phi_{B} \beta | T | \psi_\alpha \rangle = P \, h_{\beta 
\alpha}(P) \,\delta^{(3)}(\vec P_{\rm cm})\,,
\label{eq:Vab}
\end{equation}
where $P$ denotes the relative momentum of the two-meson state.

Using Eq.~(\ref{eq:funonda}) and the transition potential in 
Eq.~(\ref{eq:Vab}), we arrive to the coupled equations
\begin{equation}
c_\alpha M_\alpha +  \sum_\beta \int h_{\alpha\beta}(P) \chi_\beta(P)P^2 dP = E
c_\alpha\,, \label{ec:Ec-Res1}
\end{equation}

\begin{equation}
\sum_{\beta}\int H_{\beta'\beta}(P',P)\chi_{\beta}(P) P^2 dP +
\sum_\alpha h_{\beta'\alpha}(P') c_\alpha = E
\chi_{\beta'}(P') \,, \label{ec:Ec-Res2}
\end{equation}

where $M_\alpha$ are the masses of the bare $c\bar{c}$ mesons and 
$H_{\beta'\beta}$ is the resonant group method (RGM) Hamiltonian for the 
two-meson states obtained from the $q\bar{q}$ interaction.


\section{$P_c^+(4380)$ and $P_c^+(4450)$ resonances}

We consider the $\bar D^{(*)}\Sigma^{(*)}$ thresholds, which are the only ones 
where a sizeable residual interaction can be expected, mainly due to pion 
exchanges.

\begin{table}[!t]
\begin{center}
\begin{tabular}{cccccc}
\hline
\hline
Molecule & $J^P$ & $I$ & $Mass (MeV/c^2)$ & Width $J/\psi p$ & Width $\bar 
D^*\Lambda_c$ \\
\hline
$\bar D\Sigma_c^*$ & $\frac{3}{2}^-$ & $\frac{1}{2}$ & 4385.0 & 10.0 & 14.7 \\
\hline
$\bar D^*\Sigma_c$ & $\frac{1}{2}^-$ & $\frac{1}{2}$ & 4458.9 & 5.3 & 63.6 \\
$\bar D^*\Sigma_c$ & $\frac{3}{2}^-$ & $\frac{1}{2}$ & 4461.3 & 0.8 & 21.2 \\
$\bar D^*\Sigma_c$ & $\frac{3}{2}^+$ & $\frac{1}{2}$ & 4462.7 & 0.2 & 6.3 \\
\hline
\hline
\end{tabular}
\caption{\label{t1} Masses of the different molecular states}
\end{center}
\end{table}

In the mass region of the $P_c(4380)^+$ one can see in Table~\ref{t1} that we 
obtain one $\bar D\Sigma^*_c$ state with $J^P=\frac{3}{2}^-$. Its mass is very 
close to the experimental one and should, in principle, be identified with 
$P_c(4380)^+$.

Referring to the channel $\bar D^*\Sigma_c$ we found three almost-degenerated states 
around M=4460 MeV/$c^2$ with $J^P=\frac{1}{2}^-$, $\frac{3}{2}^-$ and 
$\frac{3}{2}^+$. The existence of these three degenerated states may be the 
origin of the uncertainty in the experimental value of $J^P$. The energy of 
those states makes them natural candidates for the $P_c(4450)^+$. It is 
remarkable that the decay width through the  $\bar D^*\Lambda_c$ channel is 
generally equal to or greater than the width via the  $J/\psi p$ channel. This 
suggests that the $\bar D^*\Lambda_c$ channel is a suitable channel for 
studying the properties of these resonances. In particular, the width of the 
predicted $\bar D^*\Sigma_c$ resonance with $J^P=\frac{1}{2}^-$ is twelve times 
greater through the $\bar D^*\Lambda_c$ channel than through the $J/\psi p$ 
channel, being this decay a good check for the existence of the resonance.

Concerning the parity of the states, a molecular scenario is not the most 
convenient to obtain positive parity states because, being the $\bar D^{(*)}$ 
mesons and the $\Sigma_c^{(*)}$ baryons of opposite parity, the relative 
angular momentum should be at least $L=1$ (P-wave) which will be above S-waves. 
This is reflected in the fact that the states with positive parity in 
Table~\ref{t1} are those with smaller binding energies.


\section{ $X(4140)$, $X(4274)$, $X(4500)$ and $X(4700)$ resonances}

Table~\ref{tab:qqbar} shows the calculated naive quark-antiquark spectrum in 
the region of interest of the LHCb for the $J^{PC}=0^{++}$ and $1^{++}$ 
channels. A tentative assignment of the theoretical states with the 
experimentally observed mesons at the LHCb experiment is also given. It can be 
seen that the naive quark model is able to reproduce all the new LHCb resonances 
except the $X(4140)$. The $X(4274)$, $X(4500)$ and $X(4700)$ appear as 
conventional charmonium states with quantum numbers $3^{3}P_{1}$, $4^{3}P_{0}$ 
and $5^{3}P_{0}$, respectively.

A complete study of the decay width of these states has been performed  
in Ref.~\cite{Ortega:2016hde} showing that the decay width of the $X(4274)$, 
$X(4500)$ and $X(4700)$ are, within errors, in reasonable agreement with the 
data. 

\begin{table}[!t]
\begin{center}
\begin{tabular}{ccccc}
\hline
\hline
State & $J^{PC}$ & $nL$  & Theory (MeV) & Experiment (MeV) \\
\hline
$\chi_{c0}$ & $0^{++}$ &  $3P$ & $4241.7$  & $-$\\
            &          &  $4P$ & $4497.2$  & $4506\pm 11^{+12}_{-15}$ \\
            &          &  $5P$ & $4697.6$  & $4704\pm 10^{+14}_{-24}$ \\
\hline
$\chi_{c1}$ & $1^{++}$ &  $3P$ & $4271.5$  &  $4273.3\pm 8.3$ \\
            &          &  $4P$ & $4520.8$  &      $-$            \\
            &          &  $5P$ & $4716.4$  &     $-$             \\           
\hline
\hline
\end{tabular}
\caption{\label{tab:qqbar} Naive quark-antiquark spectrum in the region of 
interest of the LHCb~\cite{Aaij:2016iza, Aaij:2016nsc} for the $0^{++}$ and 
$1^{++}$ channels.}
\end{center}
\end{table}

To gain some insight into the nature of the $X(4140)$, that does not appear as 
a quark-antiquark state, and to see how the coupling with the open-flavour 
thresholds can modify the properties of the naive quark-antiquark states 
predicted above, we have performed a coupled-channel calculation including the  
$D_{s}D_{s}^{\ast}$, $D_{s}^{\ast}D_{s}^{\ast}$ and $J/\psi\phi$ ones for the 
$J^{PC}=1^{++}$ sector. We found only one state with mass $4242.4\,{\rm MeV}$ 
and total decay width $25.9\,{\rm MeV}$. This state is made by $48.7\%$ of the 
$3P$ charmonium state and by $43.5\%$ of the $D_{s}D_{s}^{\ast}$ component (see 
Tables~\ref{tab:r3} and~\ref{tab:r4}). When coupling with thresholds, the 
modification in the mass is small. As we do not find any signal for the 
$X(4140)$, neither bound nor virtual, we analyse the line shape of the 
$J/\psi\phi$ channel as an attempt to explain the $X(4140)$ as a simple 
threshold cusp (see Ref.~\cite{Ortega:2016hde} for the details).

\begin{figure}[!t]
\begin{center}
\includegraphics[width=0.65\textwidth,angle=-90]{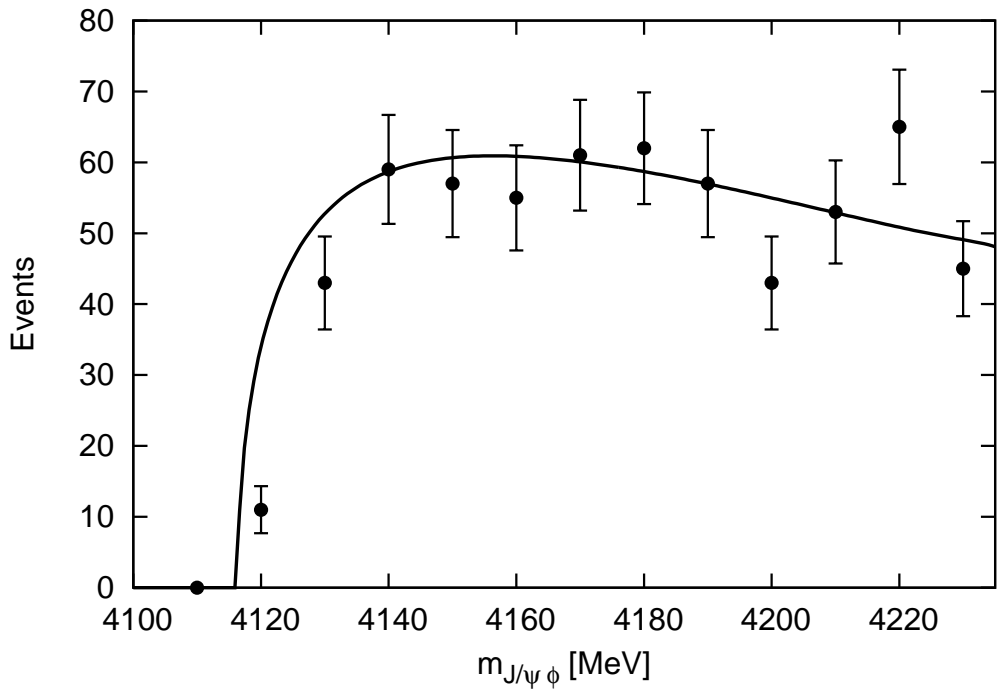}
\caption{\label{fig:f2} Line-shape prediction of the $J/\psi\phi$ channel. 
The curve shows the production of $J/\psi\phi$ pairs via direct generation from 
a point-like source plus the production via intermediate $c\bar c$ states. Note 
that the production constant has been fitted to the data (see 
Ref~\cite{Ortega:2016hde} for the details).}
\end{center}
\end{figure}

Figure~\ref{fig:f2} compares our result with that reported by the LHCb 
Collaboration in the $B^+\to J/\psi\phi K^+$ decays. The rapid increase 
observed in the data near the $J/\psi\phi$ threshold corresponds to a bump in 
the theoretical result just above such threshold. This cusp is too wide to be 
produced by a bound or virtual state below the $J/\psi\phi$ threshold. 

\begin{table}[!t]
\begin{center}
\begin{tabular}{cccccc}
\hline
\hline
Mass & Width & ${\cal P}_{c\bar c}$ & ${\cal P}_{D_{s}D_{s}^{\ast}}$ & ${\cal 
P}_{D_{s}^{\ast}D_{s}^{\ast}}$ & ${\cal P}_{J/\psi\phi}$ \\
\hline
$4242.4$ & $25.9$ & $48.7$ & $43.5$ & $5.0$ & $2.7$ \\
\hline
\hline
\end{tabular}
\caption{\label{tab:r3} Mass, total width (in MeV), and $c\bar c$ component 
probabilities (in \%) for the $X(4274)$ meson, obtained from the coupled channel 
calculation described in the text.}
\end{center}
\end{table}
\begin{table}[h]
\begin{center}
\begin{tabular}{ccccccc}
\hline
\hline
Mass (MeV) & ${\cal P}_{c\bar c}$ & ${\cal P}_{1P}$ & ${\cal P}_{2P}$ & 
${\cal P}_{3P}$ & ${\cal P}_{4P}$ & ${\cal P}_{(n>4)P}$ \\
\hline
$4242.4$ & $48.7$ & $0.000$ & $0.370$  & $99.037$ & $0.488$ & $0.105$ \\
\hline
\hline
\end{tabular}
\caption{\label{tab:r4} Probabilities, in \%, of $nP$ $c\bar c$ components in 
the total wave function of the $X(4274)$ meson.}
\end{center}
\end{table}

\section{Summary}

As a summary, our results confirm the fact that there are several states with a 
$\bar D^{(*)}\Sigma_c^{(*)}$ structure in the vicinity of the masses of the 
$P_c(4380)^+$ and $P_c(4450)^+$ pentaquark states reported by the LHCb.

Concerning the other resonances, three of them, namely $X(4274)$, $X(4500)$ and 
$X(4700)$, are consistent with bare quark-antiquark states with quantum numbers 
$J^{PC}=1^{++} (3P)$, $J^{PC}=0^{++} (4P)$ and $J^{PC}=0^{++} (5P)$, 
respectively.

In the $1^{++}$ sector we do not find any pole in the mass region of the 
$X(4140)$, though. However, the scattering amplitude shows a bump just above 
the $J/\psi\phi$ threshold which reproduces the fast increase of the 
experimental data. Therefore, the structure showed by this data around 
$4140\,{\rm MeV}$ should be interpreted as a cusp due to the presence of the 
$D_{s}D_{s}^{\ast}$ threshold.

\end{document}